\documentclass[a4paper]{spie}  

\usepackage{amsmath,amsfonts,amssymb}
\usepackage{graphicx}
\usepackage[colorlinks=true, allcolors=blue]{hyperref}

\usepackage{eurosym}

\title{First light with HiPERCAM on the GTC}

\author[a,b]{Vikram Dhillon}
\author[a]{Simon Dixon}
\author[a]{Trevor Gamble}
\author[a]{Paul Kerry}
\author[a]{Stuart Littlefair}
\author[a]{Steven Parsons}
\author[c]{Thomas Marsh}
\author[d]{Naidu Bezawada}
\author[e]{Martin Black}
\author[e]{Xiaofeng Gao}
\author[e]{David Henry}
\author[e]{David Lunney}
\author[e]{Christopher Miller}
\author[f]{Marc Dubbeldam}
\author[f]{Timothy Morris}
\author[f]{James Osborn}
\author[f]{Richard Wilson}
\author[b,g]{Jorge Casares}
\author[b,g]{Teo Mu\~{n}oz-Darias}
\author[b,g]{Enric Pall\'{e}}
\author[b,g]{Pablo Rodriguez-Gil}
\author[b,g]{Tariq Shahbaz}
\author[h]{Antonio de Ugarte Postigo}

\affil[a]{Department of Physics and Astronomy, University of
  Sheffield, Sheffield S3 7RH, UK}
\affil[b]{Instituto de Astrof\'{i}sica de Canarias, E-38205 La Laguna,
  Tenerife, Spain}
\affil[c]{Department of Physics, University of Warwick, Coventry CV4
  7AL, UK}
\affil[d]{European Southern Observatory, 85748 Garching bei M\"{u}nchen, Germany}
\affil[e]{UK Astronomy Technology Centre, Royal Observatory Edinburgh,
  Edinburgh EH9 3HJ, UK}
\affil[f]{Department of Physics, University of Durham, Durham DH1 3LE, UK}
\affil[g]{Departamento de Astrof\'{i}sica, Universidad de La Laguna, E-38206 La Laguna, Tenerife, Spain}
\affil[h]{Instituto de Astrof\'{i}sica de Andaluc\'{i}a (IAA-CSIC), E-18008, Granada, Spain}

\authorinfo{Send correspondence to VSD -- email:
  vik.dhillon@sheffield.ac.uk, tel: +44 114 222 4528}

\begin{document} 
\maketitle

\begin{abstract}
HiPERCAM is a quintuple-beam imager that saw first light on the 4.2\,m
William Herschel Telescope (WHT) in October 2017 and on the 10.4\,m
Gran Telescopio Canarias (GTC) in February 2018. The instrument uses
re-imaging optics and 4 dichroic beamsplitters to record $ugriz$
(300--1000\,nm) images simultaneously on its five CCD cameras. The
detectors in HiPERCAM are frame-transfer devices cooled
thermo-electrically to $-90^{\circ}$C, thereby allowing both
long-exposure, deep imaging of faint targets, as well as high-speed
(over 1000 windowed frames per second) imaging of rapidly varying
targets. In this paper, we report on the as-built design of HiPERCAM,
its first-light performance on the GTC, and some of the planned future
enhancements.
\end{abstract}

\keywords{instrumentation: detectors -- instrumentation: photometers
  -- techniques: photometric}

\section{INTRODUCTION}

The study of astrophysical objects that vary in brightness is set to
be revolutionised in the coming years with the advent of major new
survey facilities like Gaia, SKA, LSST and Euclid. The plethora of
interesting new targets that these facilities are expected to reveal
will require detailed follow-up on large-aperture telescopes. HiPERCAM
has been designed to perform this follow-up role, providing CCD
imaging in five optical bands ($ugriz$) simultaneously at frame rates
ranging from 0.001\,Hz to 1000\,Hz. The scientific motivation and
preliminary design of HiPERCAM have been described
before\cite{2016SPIE.9908E..0YD}, and details of the HiPERCAM data
acquisition system are described elsewhere at this
conference\cite{2018SPIE}. In this paper, we report on the final
design and first-light performance of the instrument when mounted on
the GTC on La Palma. We also describe some of the instrument
enhancements we are planning to implement in the near future.

\section{FINAL DESIGN}
\label{sec:design}

The HiPERCAM project began in January 2014, the start date of the
\euro3.5M European Research Council Advanced Grant that funded the
instrument. First light took place just under 4 years after this date,
on budget and on time.

The design of HiPERCAM is based on our successful predecessor
instrument, ULTRACAM\cite{2007MNRAS.378..825D}, but offers a very
significant advance in performance, as shown in Table~\ref{tab:comp}
and described in the following sections.

\begin{table}[htb]
\caption{Comparison of ULTRACAM and HiPERCAM.}
\label{tab:comp}
\begin{center}
\begin{tabular}{l|l|l|l}
  & {\bf ULTRACAM} & {\bf HiPERCAM} & {\bf Notes} \\
  \hline
  Number of simultaneous bands & 3 ($ug+r/i/z$) & 5 ($ugriz$) & \\
  Readout noise & 3\,e$^-$ at 100\,kHz & 4.5\,e$^-$ at 263\,kHz & Dummy output (HiPERCAM)\\
  CCD operating temperature & 233\,K & 183\,K & \\
  Dark current & 360\,e$^-$/pix/hr & 100\,e$^-$/pix/hr & \\
  Longest exposure time & 30\,s & 1800\,s & \\
  Highest frame rate & 400\,Hz & 1050\,Hz & 24$\times$24 pixel windows, bin 6$\times$6 \\
  Field of view on WHT & 5.1'$\times$5.1' & 10.2'x5.1' & platescale 0.3''/pixel \\
  Field of view on GTC & -- & 2.8'$\times$1.4' & 0.081''/pixel (HiPERCAM) \\
  Probability of $r=10$ comparison & 45\% & 78\% & WHT, galactic latitude = 30$^{\circ}$ \\
  Comparison star pick-off & No & Yes & Under development -- Sect.~\ref{sec:future} \\
  Dummy CCD outputs & No & Yes & \\
  Deep depletion & No & Yes & \\
  QE at 700/800/900/1000\,nm & 83/61/29/5\,\% & 88/78/53/13\,\% & \\
  Fringe suppression CCDs & No & Yes & \\
  Fringe amplitude in $z$ & $>$10\% & $<$1\% & \\
\end{tabular}
\end{center}
\end{table}

\subsection{Optics}
\label{sec:optics}

A ray trace through the HiPERCAM optics has been presented
before\cite{2016SPIE.9908E..0YD}. Light from the telescope is first
collimated and then split into five beams using four dichroic
beamsplitters. The beam in each arm then passes through a re-imaging
camera, which focuses the light through a bandpass filter and cryostat
window onto a CCD. The HiPERCAM collimator has been designed for use
on both the 4.2\,m William Herschel Telescope (WHT) on La Palma and
the 3.5\,m New Technology Telescope (NTT) on La Silla, giving a
platescale of 0.3''/pixel and 0.35''/pixel, and a field
of view of 11.4' and 13.4', respectively, along the diagonal of the
detector. The same collimator also gives excellent optical performance
on the 10.4\,m Gran Telescopio Canarias (GTC) on La Palma, giving a
platescale of 0.081''/pixel and a field of view of 3.1' (diagonal).

\begin{figure}[htb]
\begin{center}
\begin{tabular}{c}
\includegraphics[width=9cm]{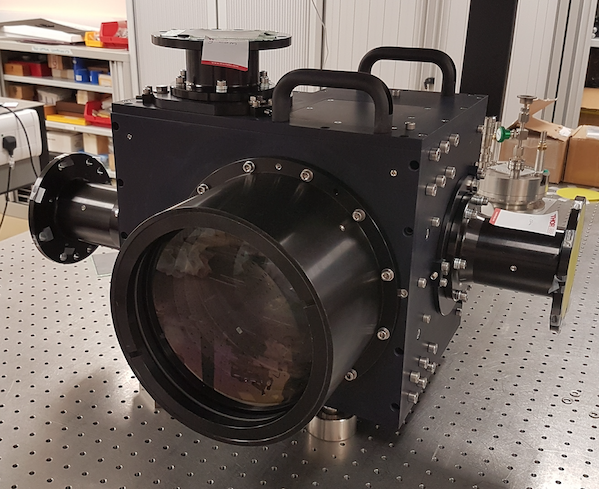}
\end{tabular}
\end{center}
\caption{The HiPERCAM hull, housing the four dichroic
  beamsplitters.  The faces of the hull act as mounting points for the
  collimator and five re-imaging cameras -- only three of the camera
  barrels are visible in this photo. The filter holders and CCD heads
  are mounted on the ends of the camera barrels. For scale, the large
  collimator lens visible in the photo has a diameter of 208 mm.
\label{fig:hull}
}
\end{figure}

\begin{figure}[htb]
\begin{center}
\begin{tabular}{c}
\includegraphics[width=10.0cm]{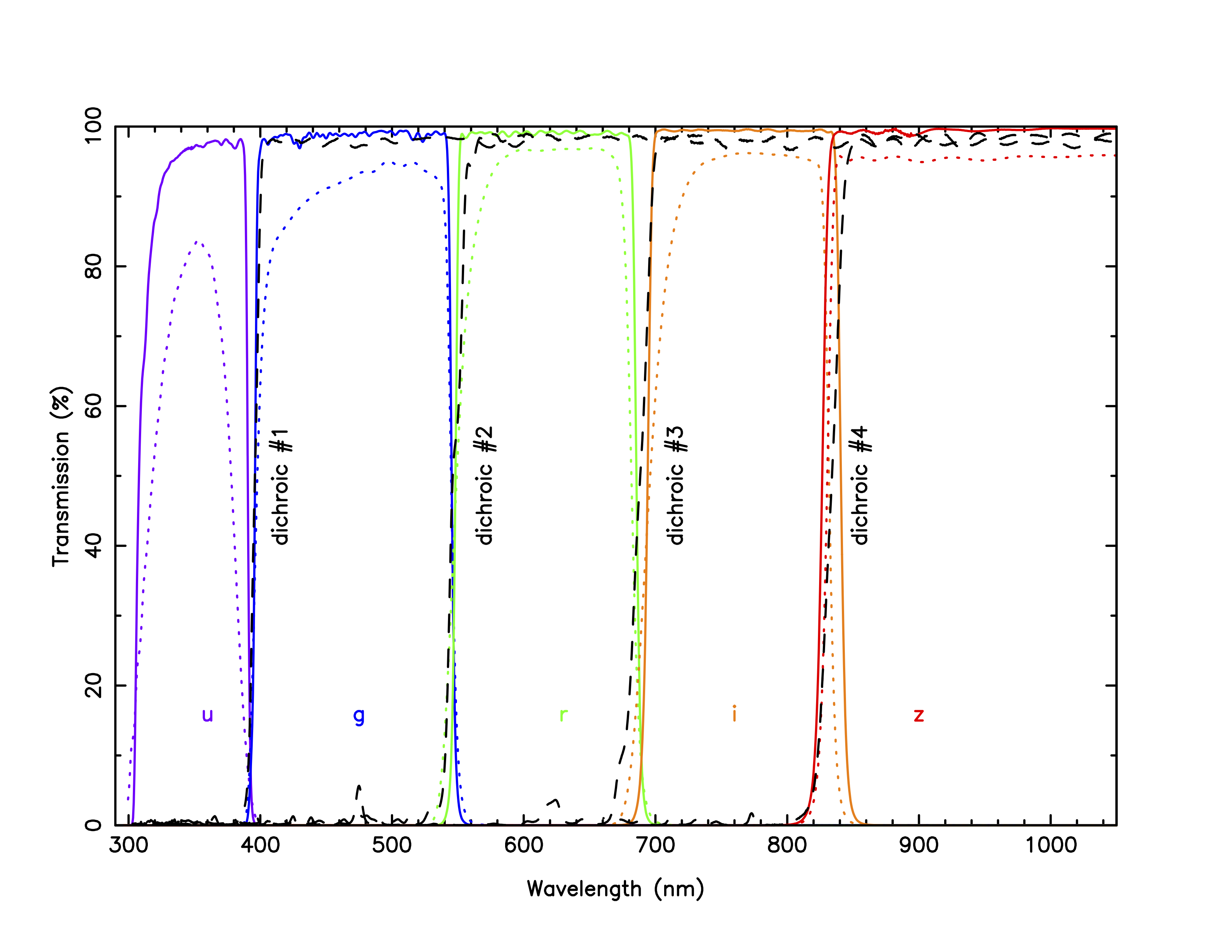}
\end{tabular}
\end{center}
\caption{As-built transmission profiles of the HiPERCAM ``Super'' SDSS
  filters (solid lines), the HiPERCAM ``original'' SDSS filters
  (dotted lines), and the four HiPERCAM dichroic beamsplitters (dashed
  lines).
\label{fig:filters}
}
\end{figure}

The dichroics, lens barrels, filters and CCDs are housed in/on an
aluminium hull, which forms a sealed system to light and dust -- see
Fig.~\ref{fig:hull}. The bandpasses of the five arms are defined by a
set of so-called ``Super'' SDSS filters (Fig.~\ref{fig:filters}),
which were designed specifically for HiPERCAM. These filters do not
use coloured glasses, but instead rely only on multi-layer coatings to
define the filter bandpasses, with the cut-on/off wavelengths designed
to match the original SDSS filter set\cite{1996AJ....111.1748F}. The
percentage improvements in throughput of the HiPERCAM Super SDSS
filters, which we call $u_sg_sr_si_sz_s$, over the original SDSS
filters, $ugriz$, are 41/9/6/9/5\,\%, respectively.

\subsection{Detectors}
\label{sec:detectors}

The HiPERCAM CCDs have been described in detail
elsewhere\cite{2016SPIE.9908E..0YD,2018SPIE}. The specifications
of the five detectors currently in use in HiPERCAM are detailed in
Table~\ref{tab:CCDs}.

\begin{table}[htb]
\caption{Specifications of the as-built HiPERCAM CCDs.}
\label{tab:CCDs}
\begin{center}
\begin{tabular}{l|l}
  \hline
  CCD model & Teledyne e2v CCD231-42 \\
  Cosmetic grade & 1 \\
  Architecture & Split frame transfer, back thinned, 2-phase, NIMO \\
  Format & 2048$\times$2048 pixels \\
  Image area & 2048$\times$1024 pixels \\
  Storage area & 2$\times$ 2048$\times$512 pixels \\
  Pixel size & 15\,$\mu$m \\
  Outputs & 4 \\
  Readout noise & 3.2\,e$^-$ at 200\,kHz, single-ended\\
  Gain & 1.2\,e$^-$/ADU \\
  Dark current & 10\,e$^-$/pix/hr at 173\,K \\
  Deep depletion Si & $i_s$ and $z_s$ bands \\
  AR coatings & Astro Broadband in $u_s$, Astro Multi-2 in $g_s r_s i_s z_s$ \\
  Peak QE in $u_sg_sr_si_sz_s$ bands & 76/89/88/88/78\,\%\\
  Fringe suppression & $i_s$ and $z_s$ bands \\
  Fringe amplitude & $\sim0.1$\% in $i_s$ and $\sim1$\% in $z_s$ band \\
  Full well & 120 ke$^-$ \\
  Non linearity & $< \pm$0.5\% \\
  Vertical clocking & 10\,$\mu$s/row \\
  Horizontal clocking & 0.1\,$\mu$s/pix, split serial-register \\
  Full frame time (bin 1$\times$1) & 3.0\,s in slow, 1.25\,s in fast readout mode \\
  Full frame time (bin 2$\times$2) & 0.9\,s in slow, 0.4\,s in fast readout mode \\
  Drift-mode frame time & 0.0009\,s with 24$\times$24 pixel windows, bin 6$\times$6 \\
\end{tabular}
\end{center}
\end{table}

\begin{figure}[htb]
\begin{center}
\begin{tabular}{c}
\includegraphics[width=7cm]{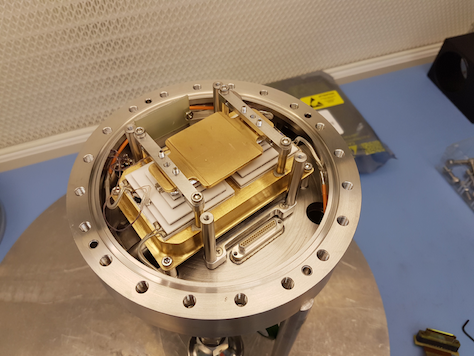}~\includegraphics[width=7cm]{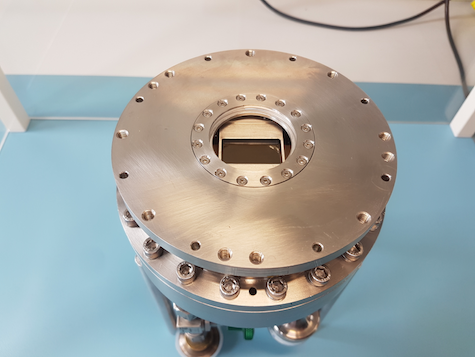}
\end{tabular}
\end{center}
\caption{Left: Interior view of one of the HiPERCAM CCD heads, showing
  the gold-plated cold plate sitting on top of two, white 5-stage
  TECs, which themselves are sitting on a gold-plated heatsink through
  which the cooling liquid runs. Right: Exterior view of one of the
  HiPERCAM CCD heads.  The diameter of the head is 160\,mm and the
  weight is approximately 7\,kg.
\label{fig:head}
}
\end{figure}

The HiPERCAM detectors need to be cooled to below 187\,K in order to
achieve the dark current requirement of $<360$\,e$^-$/pixel/hr, which
corresponds to 10\%\ of the faintest sky level we can observe with
HiPERCAM (set by observations in the $u$-band in dark time on the
GTC). Cooling to below 187\,K therefore ensures that dark current is
always a negligible noise source in HiPERCAM. We looked at a number of
cooling options\cite{2016SPIE.9908E..0YD} before deciding to use
thermo-electric (peltier) coolers (TECs), which are the simplest,
cheapest, lightest and most compact of coolers. Our solution uses two
Marlow NL5010 five-stage TECs mounted side by side, as shown in
Fig.~\ref{fig:head}. Our detector head design uses all-metal seals
rather than o-rings in order to minimize leaks. We went to great
lengths to avoid using any materials inside the detector heads that
could outgas. So, for example, we mounted the pre-amplifier board
outside the head\cite{2018SPIE}, and used a corrugated indium foil to
make good thermal connections between the heatsink, TECs and cold
plate. We also reduced outgassing by thoroughly cleaning all
components prior to assembly, and then baking the assembled head
whilst vacuum pumping. Even with all of these precautions, outgassing
limits the vacuum hold time of the HiPERCAM CCD heads to around 1
week, due primarily to the small interior volume of the heads
($\sim0.5$\,litres) and the lack of a sufficiently cold, large-area
interior surface to give effective cryopumping. Fortunately, it only
takes a few minutes to pump down the heads whilst on the telescope,
thanks to the use of a 5-way vacuum manifold system permanently
installed on the instrument.

The heat generated by the TECs is extracted using a 278\,K
water-glycol cooling circuit. To ensure that cooling fluid of the same
temperature enters each of the 5 CCD heads, the heads are connected in
parallel rather than series, via two 6-way manifolds (the sixth arm is
for cooling the NGC controller). Each arm in this parallel circuit is
equipped with a flow sensor connected to a Honeywell Minitrend GR Data
Recorder mounted in the electronics cabinet. As well as providing a
display of the flow rate through each CCD head, the data recorder has
relays that can switch off the power to the TEC power supplies if the
flow rate in any head drops below a user-defined limit, thereby
protecting the CCDs from overheating. As a backup to this system, the
TEC power supplies themselves (made by Meerstetter, model LTR-1200)
have a high-temperature cut-off facility: if the temperature of the
heat-sink in the CCD head rises above a user-defined value, such as
would occur if the coolant supply fails, the power to the TEC is
automatically shut off. The TEC power supplies are able to maintain
the CCD temperatures at their $-90^{\circ}$C set points to within
$0.01-0.1^{\circ}$C.

In case of high humidity, HiPERCAM has a 5-way manifold that enables
clean, dry air from a telescope supply to be blown across each of the
CCD windows at approximately 1\,litre/min to prevent condensation.

\subsection{Mechanics}
\label{sec:mechanics}

\begin{figure}[htb]
\begin{center}
\begin{tabular}{c}
\includegraphics[width=10.0cm]{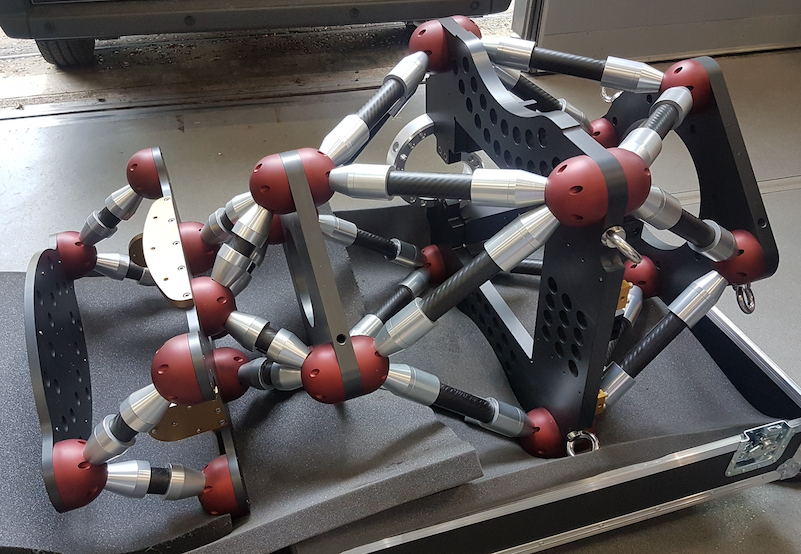}
\end{tabular}
\end{center}
\caption{The empty HiPERCAM opto-mechanical chassis sitting in its
  packing crate. The plate at the right-hand side mounts onto the
  telescope. The large central plate with a square aperture in it
  houses the hull. The CCD controller is mounted between the two
  left-most plates.
\label{fig:octopod}
}
\end{figure} 

The HiPERCAM opto-mechanical chassis is a triple octopod composed of 3
large aluminium plates connected by carbon fibre struts, as shown in
Fig.~\ref{fig:octopod}. This design ensures a stiff, compact (1.25\,m
long), light-weight (220\,kg) and open structure, which is relatively
insensitive to temperature variations. These characteristics make
HiPERCAM easy to maintain, transport and mount/dismount at the
telescope.

\begin{figure}[htb]
\begin{center}
\begin{tabular}{c}
\includegraphics[width=6cm]{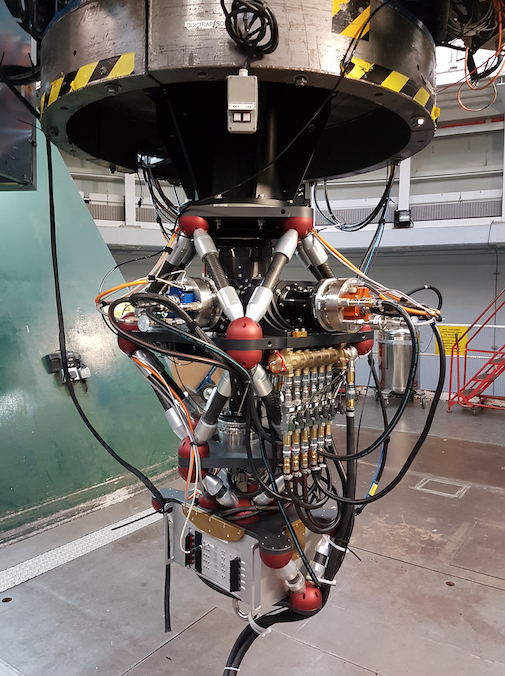}~\includegraphics[width=9cm]{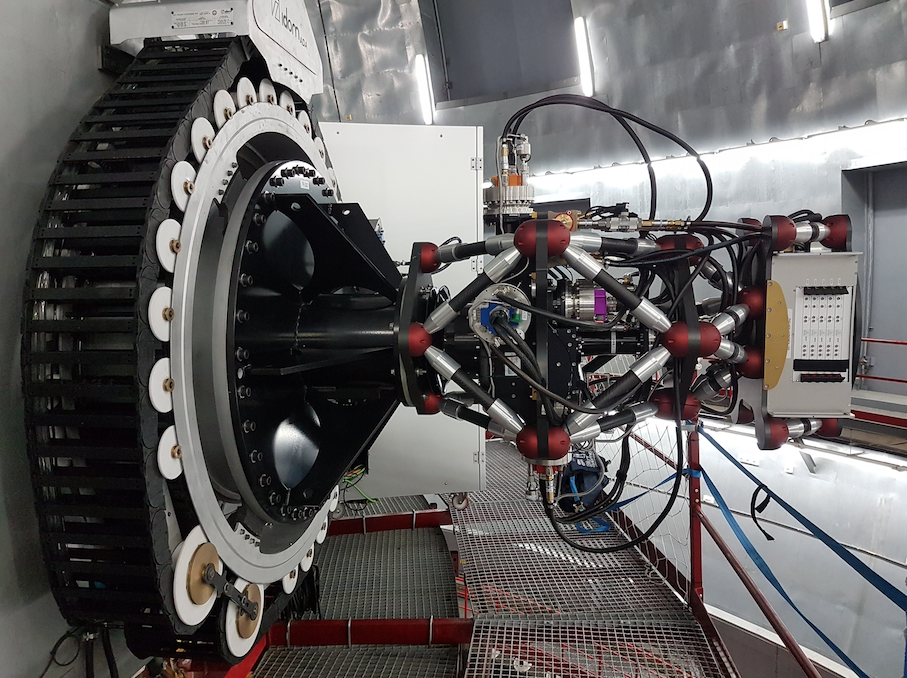}
\end{tabular}
\end{center}
\caption{Left: HiPERCAM mounted at the Cassegrain focus of the WHT. Right:
  HiPERCAM mounted on the Folded Cassegrain focus of the GTC.
\label{fig:telescope}
}
\end{figure}

The top plate of the triple octopod is used to mount the instrument
onto the telescope, the middle plate is used to mount the hull on
(Fig.~\ref{fig:hull}), and the bottom plate is used to mount the CCD
controller onto the instrument. Two different interface collars are
used to attach HiPERCAM onto the rotator of the WHT and GTC, and place
the instrument at the correct distance from the respective telescope
focal planes (see Fig.~\ref{fig:telescope}).

\subsection{Data acquisition system}
\label{sec:das}

The HiPERCAM data acquisition system (DAS) has been described in
detail elsewhere\cite{2016SPIE.9908E..0YD,2018SPIE}. The DAS in
HiPERCAM is detector limited, i.e. the throughput of data from the
output of the CCDs to the hard disk on which it is archived is always
greater than the rate at which the data comes off the CCDs. This means
that the instrument is capable of running continuously all night at
its maximum data rate without ever having to pause for archiving of
data. All CCDs are read out simultaneously and have identical exposure
start and end times. In order to change the exposure times of the CCDs
with respect to each other, it is possible to skip the readout of
selected CCDs using the {\em NSKIP} parameter. For example, if NSKIP
is set to 3,2,1,2,3 for the $u,g,r,i,z$ CCDs, and the exposure time is
set to 10\,s, then the CCD controller will read out only the $r$-band
CCD on the first readout cycle (giving $r$ a 10\,s exposure), then the
$g$, $r$ and $i$-band CCDs on the second cycle (giving $g$ and $i$ a
20\,s exposure), then the $u$, $r$ and $z$-band CCDs on the third
cycle (giving $u$ and $z$ a 30\,s exposure), etc.

\begin{figure}[htb]
\begin{center}
\begin{tabular}{c}
\includegraphics[width=10.0cm]{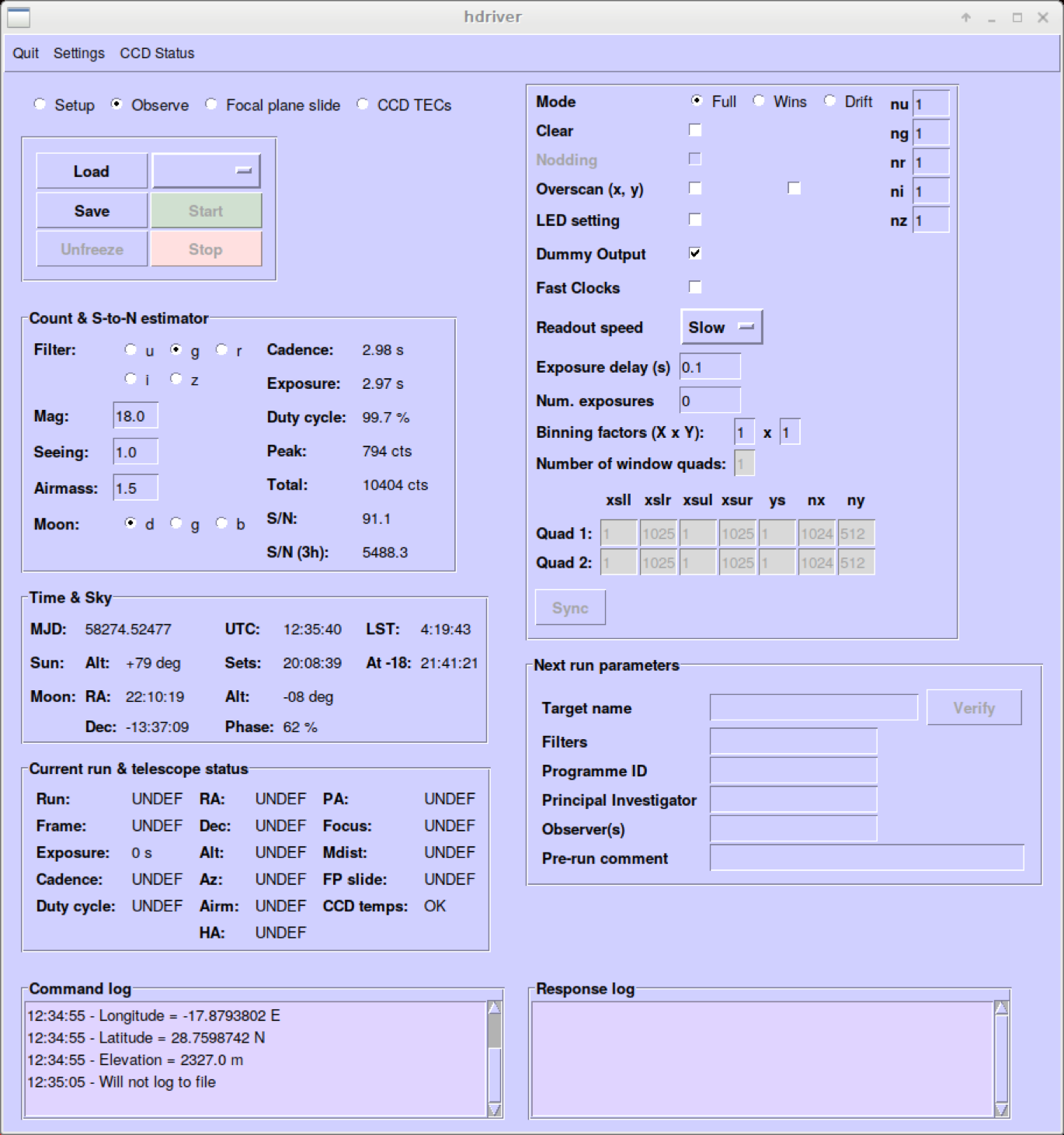}
\end{tabular}
\end{center}
\caption{A screenshot of the python-based HiPERCAM GUI, used by
  astronomers at the telescope to control the instrument. There also
  exists an engineering GUI for low-level control and telemetry of the
  CCD controller.
\label{fig:gui}
}
\end{figure} 

\begin{figure}[htb]
\begin{center}
\begin{tabular}{c}
\includegraphics[width=12.0cm]{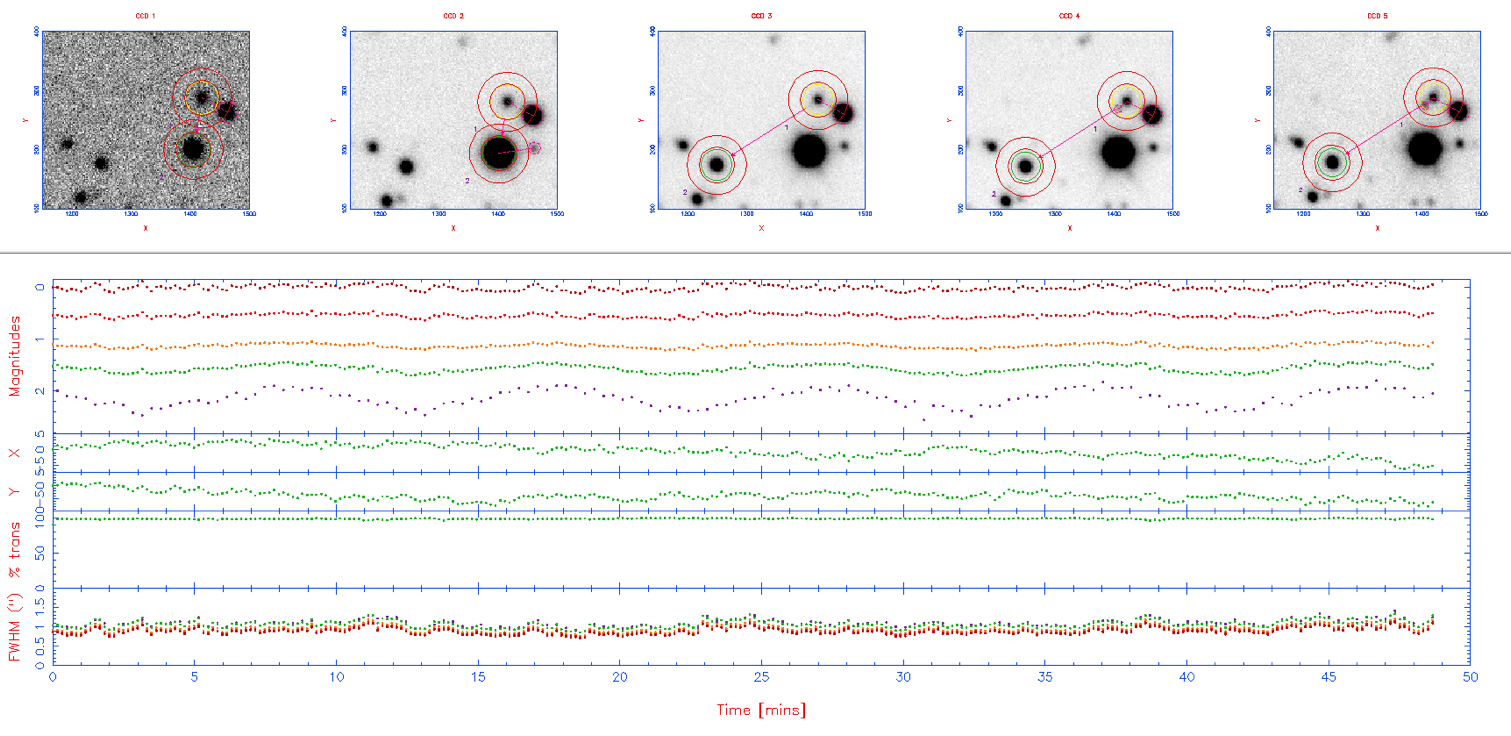}
\end{tabular}
\end{center}
\caption{A screenshot of the python-based HiPERCAM data reduction
  pipeline. The top row shows a zoom-in of the $u_sg_sr_si_sz_s$
  images of the target and comparison stars, with the apertures
  defining the object and sky regions superimposed. The bottom panel
  shows the target minus comparison star magnitudes in
  $u_sg_sr_si_sz_s$ (top row), the comparison star $x,y$ positions
  (second and third rows), the sky transparency measured from the
  comparison star flux (fourth row), and the seeing in
  $u_sg_sr_si_sz_s$ measured from the comparison-star FWHM (bottom
  row).
\label{fig:pipeline}
}
\end{figure}

The HiPERCAM graphical user interface (GUI) is shown in
Fig.~\ref{fig:gui}. The top right of the GUI shows three buttons that
allows the astronomer to change the readout mode: full frame, windowed
or drift mode\cite{2007MNRAS.378..825D}. Below this are the various
CCD readout parameters, giving the astronomer complete control over
the detector setup. The number of exposures is usually set to zero in the GUI
and the green {\em Start} button is then pressed: the data acquisition
system will then take data continuously until the red {\em Stop}
button is pressed. All frames of a HiPERCAM run on a target are
written to a single, custom-format FITS file. The HiPERCAM
data-reduction pipeline software (see Fig.~\ref{fig:pipeline}) reads
this file and provides a quick-look reduction of the data whilst
observing.  The pipeline is a fully-featured photometry reduction
package and so can also be used off-line to produce publication-ready
light curves.

\subsection{On-sky performance}
\label{sec:performance}

We commissioned HiPERCAM on the sky for the first time on the WHT
during October 2018. This was a short run of only 1 commissioning
night and 4 science nights, and was intended primarily to be a
shake-down of the instrument prior to commissioning on the GTC. The
instrument was then moved to the GTC, where we were allocated 3
commissioning nights and 10 science nights in February 2018. We
suffered from terrible weather during this run, and only observed for
the last 3 nights, when we completed all of the commissioning.  Three
more observing runs followed in April, May and June 2018, totalling 16
nights. A wide range of science was performed during this time,
including the observation of black holes, white dwarfs, neutron stars,
brown dwarfs, extrasolar planets/asteroids, AGN, FRBs, GRBs, SNe and
ultra-diffuse galaxies. We report on the performance of HiPERCAM on
the GTC below.

\begin{figure}[htb]
\begin{center}
\begin{tabular}{c}
\includegraphics[width=12.0cm]{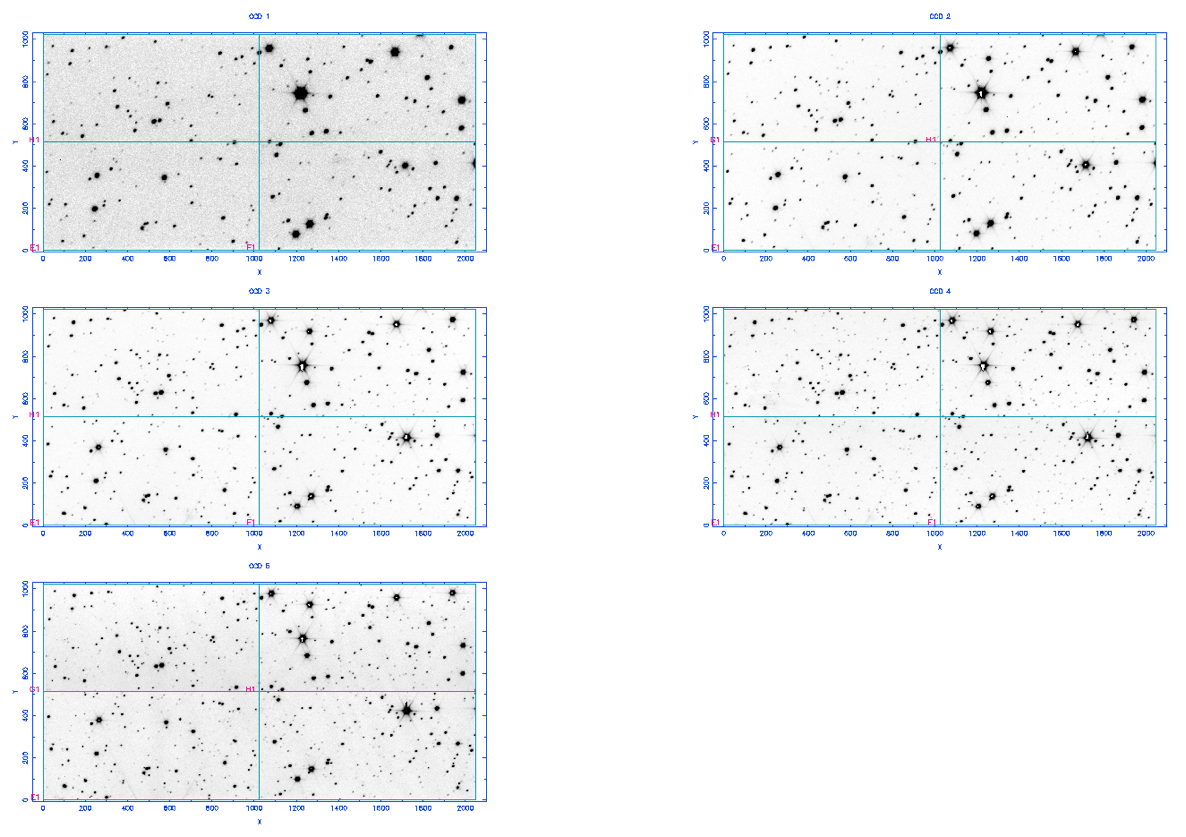}
\end{tabular}
\end{center}
\caption{Images of a star field obtained with HiPERCAM on the GTC in
  $u_s$ and $g_s$ (top row), $r_s$ and $i_s$ (middle row), and $z_s$
  (bottom row).
\label{fig:starfield}
}
\end{figure}

An example star field observed with HiPERCAM on the GTC is shown in
Fig.~\ref{fig:starfield}. The FWHM of the stars in these images are
$u_s = 0.56''$, $g_s = 0.44''$, $r_s = 0.41''$, $i_s = 0.37''$, $z_s =
0.36''$, with no discernible variation with field angle. This
indicates that HiPERCAM on the GTC can provide seeing-limited images
across the whole field of view in even the very best seeing conditions
on La Palma. Since the dichroics operate in a collimated beam and
have anti-reflection coatings on their rear surfaces, we do not expect
to see any ghosting in our images, and this is indeed the case: a
careful inspection of the brightest stars in Fig.~\ref{fig:starfield}
reveals no discernible ghosting.  The pixel positions of the stars at
the corners of the field of view are the same on all 5 CCDs, to within
approximately 5 pixels (75\,$\mu$m), indicating that there is no
discernible variation of platescale with wavelength and that the CCD
heads are well aligned with respect to each other. Twilight-sky flat fields
show no discernible vignetting in the corners of the field of view.

The measured photometric zero points of HiPERCAM on the GTC, defined
here as the magnitude of a star that would give 1 electron per second
above the atmosphere, are: $u_s = 26.92~(25.76)$, $g_s = 28.77~
(28.26)$, $r_s = 28.54~(28.84)$, $i_s = 28.30~(28.49)$, $z_s = 27.85~
(27.95)$. The numbers in brackets show the corresponding values for
OSIRIS, the common-user, single-channel optical imager at the
GTC\cite{2003SPIE.4841.1739C}. These two sets of zero points were
measured within a few days of each other during May 2018, using
observations of SDSS standard stars. It can be seen that unless one
needs the full 7.8' field of view of OSIRIS, or just one of the redder
bands, it is much more efficient to use HiPERCAM for standard imaging
at the GTC, and one would also gain from the huge reduction in dead
time between exposures (0.01\,s with HiPERCAM, 21\,s with OSIRIS).

\begin{figure}[htb]
\begin{center}
\begin{tabular}{c}
\includegraphics[width=10.0cm]{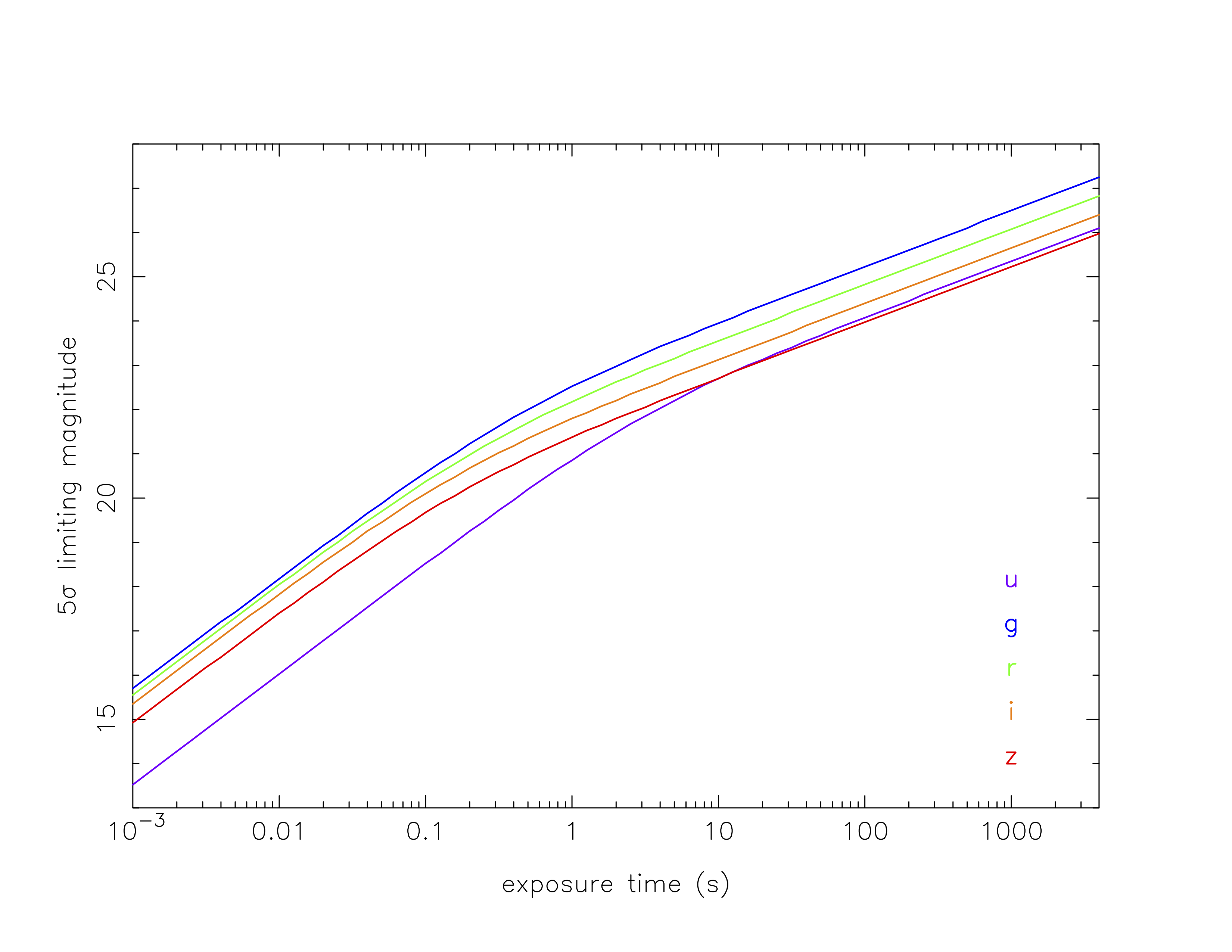}
\end{tabular}
\end{center}
\caption{Limiting magnitudes (5$\sigma$) of HiPERCAM on the GTC as a
  function of exposure time. The purple, blue, green, orange and red
  curves show the results for the $u_sg_sr_si_sz_s$
  filters, respectively. The calculations assume dark moon,
  observing at the zenith and seeing of 0.8''.
\label{fig:limitingmags}
}
\end{figure}

The 5$\sigma$ limiting magnitudes of HiPERCAM on the GTC are shown as
a function of exposure time in Fig.~\ref{fig:limitingmags}, calculated
using the zero points given above. It can be seen that it is possible
to achieve a limiting magnitude of $g\sim 16$ in a single 0.001\,s
exposure, $g\sim 23$ in 1\,s, and $g\sim 27$ in 1800\,s.

\begin{figure}[htb]
\begin{center}
\begin{tabular}{c}
\includegraphics[width=10.0cm]{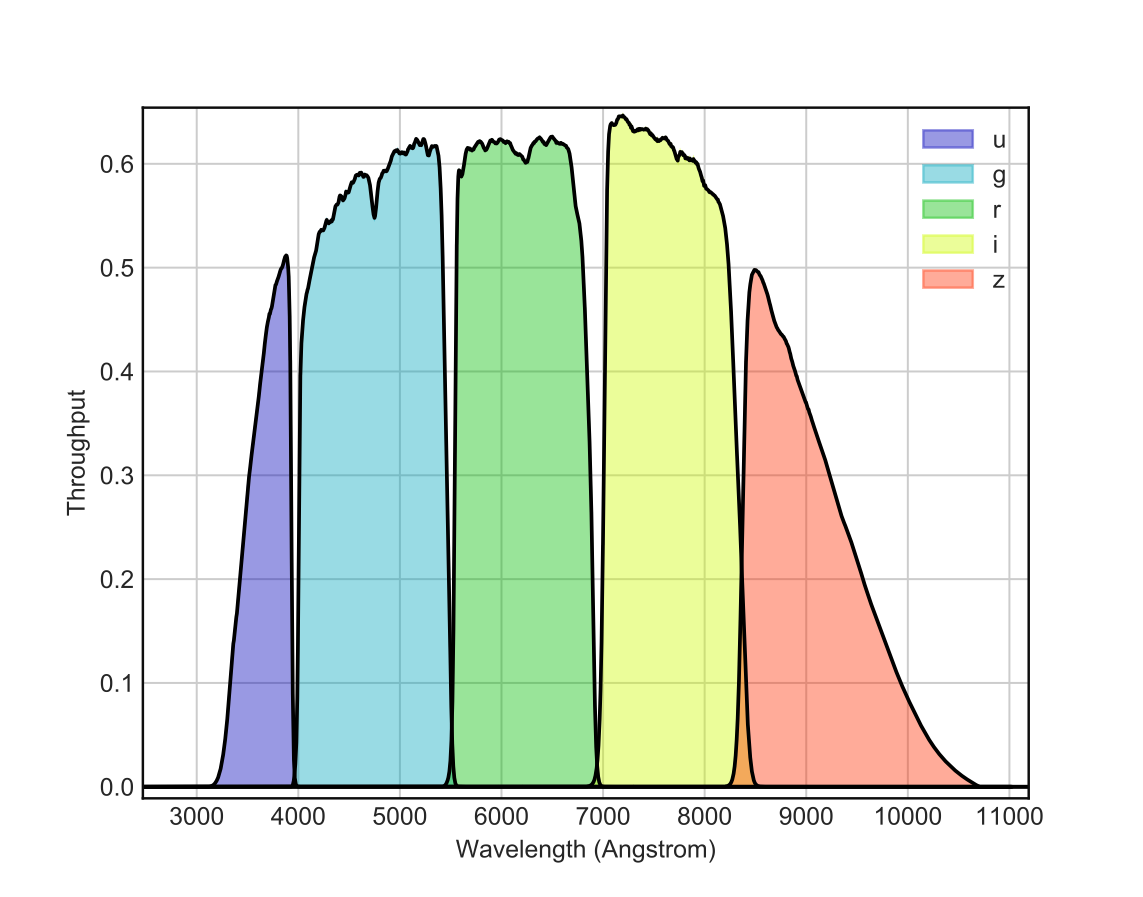}
\end{tabular}
\end{center}
\caption{Throughput curves for HiPERCAM in $u_s g_s r_s i_s
  z_s$. These curves do not include the atmosphere or telescope.
\label{fig:throughput}
}
\end{figure}

The throughput curves of HiPERCAM, which include all optics and CCDs
but not the atmosphere and telescope, are shown in
Fig.~\ref{fig:throughput}. The throughput peaks at over 60\% in $g_s$,
$r_s$, $i_s$ and 50\% in $u_s$, $z_s$. This high throughput has been
achieved by using high-performance multi-layer coatings on the lenses,
dichroics, filters and windows, as well as CCDs optimised for
operation in each band.

\begin{figure}[htb]
\begin{center}
\begin{tabular}{c}
\includegraphics[width=10.0cm]{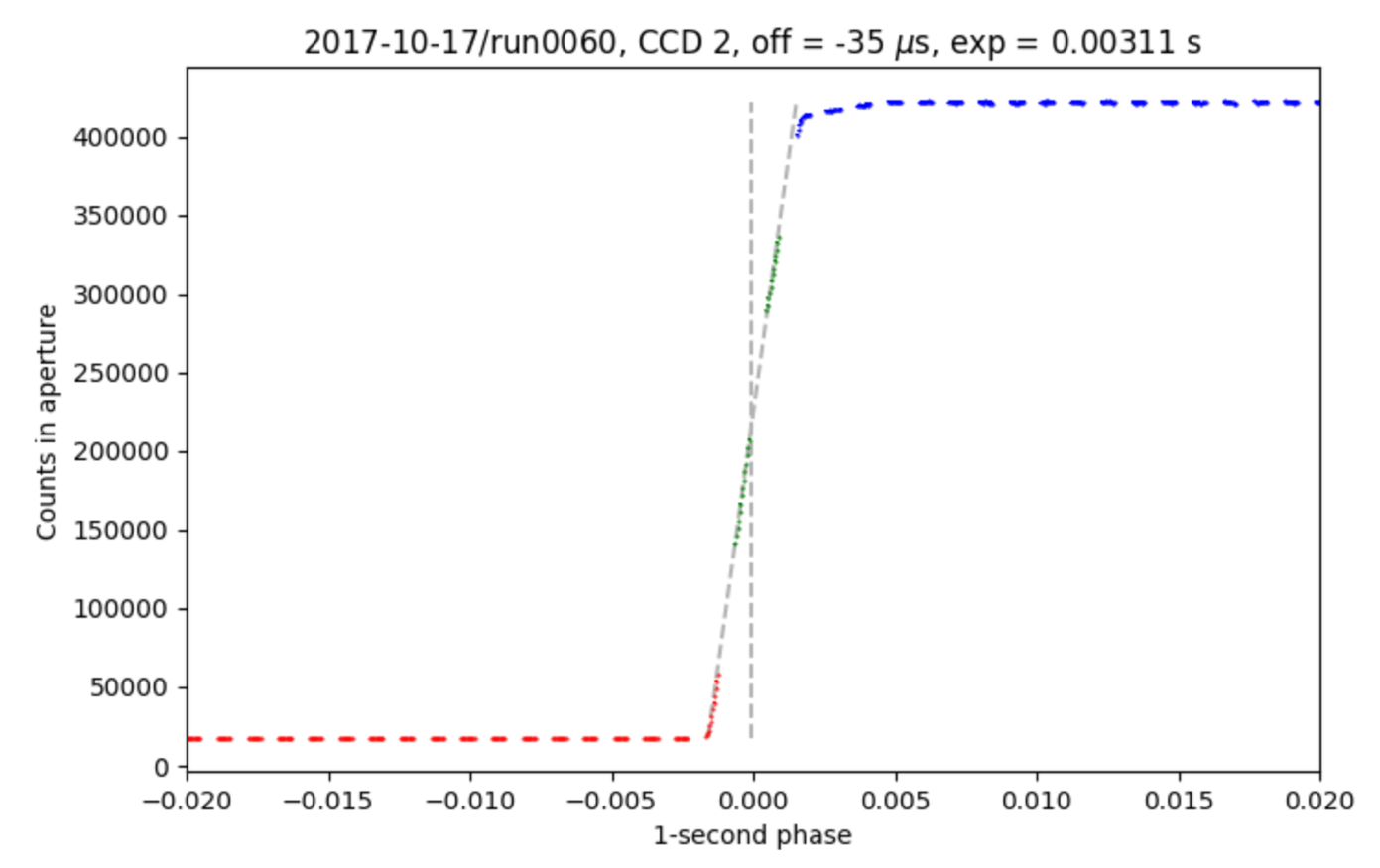}
\end{tabular}
\end{center}
\caption{Phase-folded light curve of an LED attached to the PPS output
  of the HiPERCAM GPS system.
\label{fig:gps}
}
\end{figure}

We measured the accuracy of the GPS absolute timestamping of each CCD
frame in HiPERCAM by observing an LED connected to the
pulse-per-second (PPS) output of our GPS system. An example
observation is shown in Fig.~\ref{fig:gps}, where we plot the light
curve of the LED phase-folded on its 1\,s period. The shape of the
light curve is a convolution of two top-hat functions, one for the LED
pulse and the other for the exposure time duration (in this example,
0.003\,s). Hence the folded light curve exhibits a ramp, the centre of
which should correspond to the start of the GPS
second. Fig.~\ref{fig:gps} shows that this is indeed the case: the
offset between the start of the GPS second and centre of the ramp is
only 35\,$\mu$s and is set by the measurement accuracy of our
experiment. Hence we can say that the absolute timestamping of each
CCD frame in HiPERCAM is accurate to better than tens of microseconds.

\subsection{Future plans}
\label{sec:future}

HiPERCAM was removed from the GTC rotator in June 2018 so that an
autoguider can be installed. The instrument will be remounted on the
telescope from September 2018 to the end of the year for more science
runs. In 2019, HiPERCAM will share the Folded Cassegrain focus of the
GTC with CanariCam\cite{2008SPIE.7014E..0RT}. The situation from 2020
onwards is less clear at the moment, due to the arrival of MIRADAS at
the GTC\cite{2016SPIE.9908E..1LE}.

We are planning on making two enhancements to HiPERCAM during the
coming year. The first is to modify the CCD preamp boards so that it
is possible to switch the bandwidth from the current value of
1.06\,MHz\cite{2018SPIE} to approximately half this value. In this way,
we hope to reduce the readout noise from its current value of $\sim
4$\,e$^-$ to $\sim 3$\,e$^-$ in slow readout mode (263 kHz).

\begin{figure}[htb]
\begin{center}
\begin{tabular}{c}
\includegraphics[width=12.0cm]{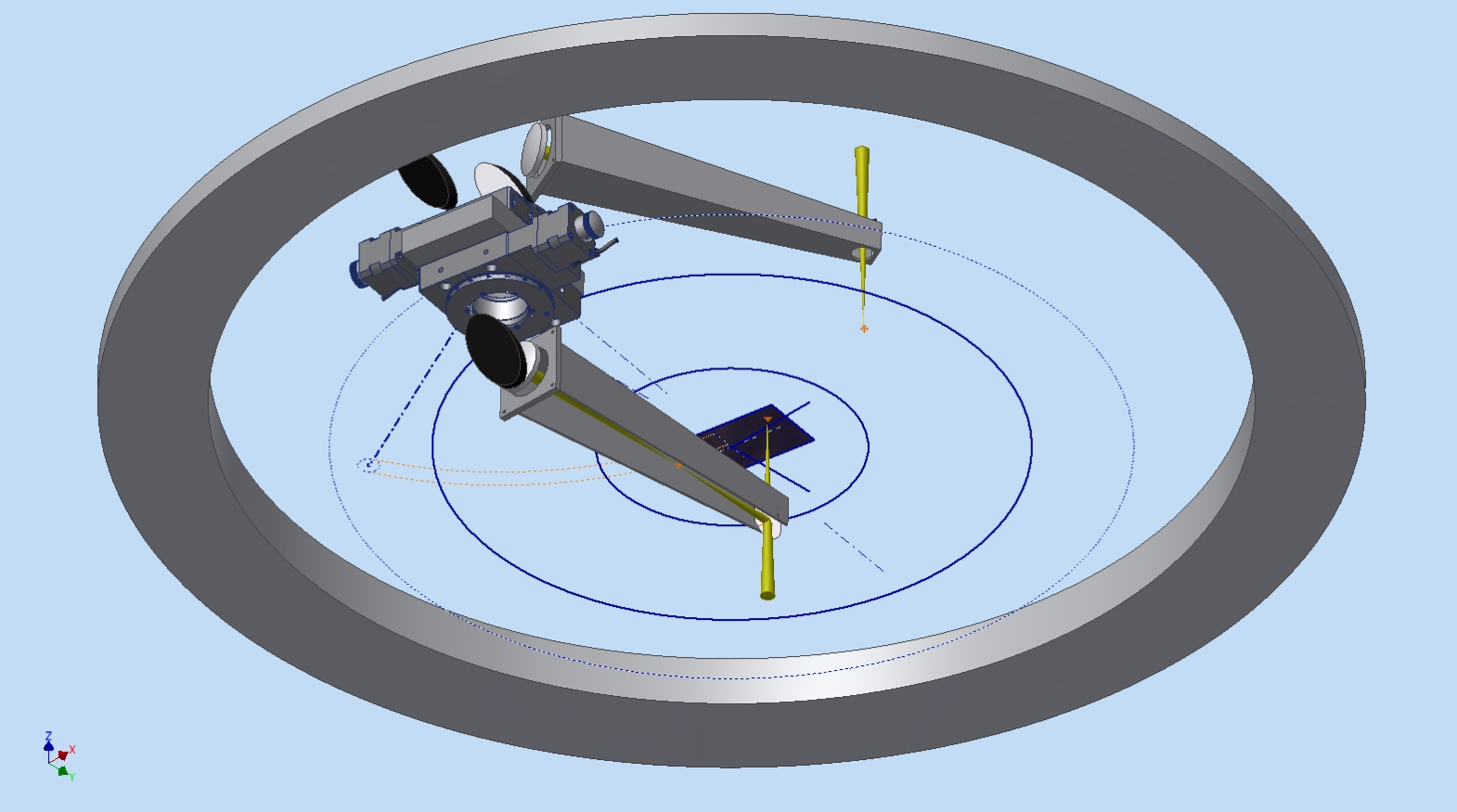}
\end{tabular}
\end{center}
\caption{CAD image of COMPO. The upper arm collects light from a star
  falling outside the HiPERCAM field of view (solid blue rectangle at
  centre) but inside the 10' diameter view of view at the Folded
  Cassegrain focus of the GTC (outer blue circle). The lower arm
  redirects this light to one of the corners of the HiPERCAM field of
  view, via a set of relay optics.
\label{fig:compo}
}
\end{figure}

The second enhancement is to implement a COMParison star Pick-Off
system (COMPO), as shown in Fig.~\ref{fig:compo}. Light from a bright
comparison star that falls outside the 3.1' diagonal field of view of
HiPERCAM, but within the 10' diameter field of view of the Folded
Cassegrain focus of the GTC, is collected by a pick-off arm. The light
is then redirected to a second arm, via a set of relay optics, which
injects the starlight onto one of the corners of the HiPERCAM CCDs. In
this way it is possible to use much brighter comparison stars than
would usually be available for differential photometry. This will be
of particular importance when observing bright targets, like exoplanet
host stars, for which nearby comparison stars of comparable or greater
brightness are rare. More quantitatively, without COMPO, there is a
90\%\ probability of finding a comparison star of magnitude $r=14$ in
the field of view, whereas with COMPO one will be able to find
comparison stars of magnitude $r=12$ with the same probability. COMPO
will also be of great benefit to most $u_s$-band observations, as many
of the targets observed by HiPERCAM tend to be blue (e.g. white
dwarfs) and hence bright in the $u_s$-band, whereas most comparison
stars tend to be red and hence faint in the $u_s$-band.

\section{CONCLUSIONS}

We have described the as-built design and first-light performance of
HiPERCAM on the GTC, as well as some of the enhancements we are
planning to implement over the coming year. The on-sky results
demonstrate that HiPERCAM is performing to specification. The
instrument is now entering its exploitation phase on the GTC, and
promises to revolutionise the field of high time-resolution optical
astrophysics.

\acknowledgments
 
HiPERCAM is funded by the European Research Council under the European
Union's Seventh Framework Programme (FP/2007-2013) under ERC-2013-ADG
Grant Agreement no. 340040 (HiPERCAM). We would like to thank the
staff of the mechanical workshops at the University of Sheffield and
the UKATC for their major contribution to the project. We would also
like to thank the staff of the ING and GTC for their assistance during
commissioning.

\bibliography{hipercam_spie2} 
\bibliographystyle{spiebib} 

\end{document}